\documentclass[runningheads]{llncs}

\usepackage{header}
\usepackage{amssymb}

\begin{document}

\title{Mixed-Precision SEM-Based CFD Simulations on GPUs: A Taylor-Green Vortex case}
\titlerunning{Mixed-Precision SEM-Based CFD Simulations on GPUs}

\author{Yanxiang Chen\inst{1}, Manuel M\"unsch\inst{2}, and Roman Iakymchuk\inst{1,3}}
\authorrunning{Y~Chen, M~M\"unsch, R~Iakymchuk}

\institute{
Ume\aa\  University, Sweden
\email{\{ychen,riakymch\}@cs.umu.se}
\and
Friedrich-Alexander University Erlangen-Nuremberg, Germany
\email{manuel.muensch@fau.de}
\and
Uppsala University, Sweden
}

\maketitle

\begin{abstract}
\vspace{-15pt}
Mixed precision is a promising approach for reducing the computational cost and energy consumption of Computation Fluid Dynamics (CFD) simulations, but its effectiveness depends strongly on where precision is reduced within the full simulation pipeline. In this work, we study Taylor–Green vortex case using Neko, a matrix-free CFD solver based on the spectral element method (SEM). Profiling shows that the fluid time step is not dominated by Krylov convergence alone: the velocity and pressure solvers require only a small number of iterations per step, while a substantial fraction of runtime is spent in other SEM operators and solver components. Motivated by this structure, we propose a three-level hierarchical mixed-precision control model. Two groups of configurations are evaluated in environments bounded by 64-bit floating-point (fp64) and 32-bit floating-point (fp32) precision, respectively. The fp64-bounded group identifies accuracy sensitive components and shows that SEM-focused fp32 computation is a promising direction for future optimization. The fp32-bounded group provides the main practical benefit. For the high Reynolds number case studied, selected configurations reduce both time- and energy-to-solution by about $34\%$ relative to the fp64 baseline, while improving robustness compared with global fp32. Targeted fp16 kernel overrides are also explored, showing potential for selected operations but increased sensitivity in gradient-based quantities such as enstrophy. Overall, these results indicate that mixed-precision for matrix-free SEM-based CFD should be treated as a simulation-level control problem rather than solely as a Krylov-solver optimization.
\vspace{-5pt}
\end{abstract}

\section{Introduction}
Traditionally, scientific applications have been developed using double precision, i.e., fp64, the 64-bit IEEE 754 floating-point format. Lower precisions such as single precision, fp32, have long been recognized for their potential to reduce both memory traffic and execution time. More recently, AI-driven even lower-precision formats, including fp16, bf16, and int8, have emerged as promising candidates for scientific computing~\cite{duben19}, but require further investigation. In this context, mixed-precision algorithms have been extensively studied in numerical linear algebra~\cite{anztsurvey,theo22}, motivated by the fact that many scientific applications require high-accuracy solutions while also seeking performance and energy gains through the use of lower precision in selected parts of the computation.

Many applications, including Computational Fluid Dynamics (CFD), ultimately rely on the solution of large linear systems $Ax=b$. In practice, iterative methods, particularly Krylov subspace methods, are among the most commonly used solvers. To balance accuracy and efficiency, numerous mixed-precision Krylov methods have been proposed, such as GMRES-based iterative refinement using multiple precisions~\cite{cahi18,gmres-ir5}. Most of these methods are matrix-based, in the sense that they assume explicit access to the stored matrix $A$ and, often, to a preconditioner $M$, as in the LU-based iterative refinement~\cite{cahi18}. Within this framework, the storage precision of the system matrix itself also becomes an important design question. Mixed-precision variants of the conjugate gradient (CG) method, e.g. with adaptive precision block-Jacobi preconditioner~\cite{Flegar2021Ginkgo}, have likewise been proposed. We refer to such approaches as fine-grained mixed-precision algorithms, since the precision-switching strategies are applied within the internal operations of a Krylov solver. 
In this paper, we demonstrate the limited applicability of fine-grained mixed-precision algorithms to SEM-based CFD applications and propose a hierarchical multi-level approach for enabling mixed precision.

Based on the previous analysis of the Poisson's example with Nekbone~\cite{ppam24} and Neko~\cite{fgcs26} with the help of Verificarlo computer arithmetic tool~\cite{Denis2016verificarlo}, where only the velocity solve of the Navier-Stokes equation was involved, we proceed with the full SEM simulation invoking both velocity and pressure solves of the Navier-Stokes equation on the example of Taylor-Green Vortex. Our main contributions are as follows:\\[-6mm]
\begin{enumerate}
    \item Proposed general methodology for enabling mixed-precision in different parts of SEM-based CFD simulation, including the fluid solver, Krylov-type solvers, and kernels. 
    \item Extensive study on the Taylor-Green Vortex benchmark: We increase not only the simulation size but also the polynomial order to gain deeper insight into the factors that affect result accuracy and determine the intermediate precision required by these solvers.
    \item Evaluated the proposed mixed-precision strategies on a few European supercomputers, focusing on GPUs with a wider range of hardware supported low precision formats.
\end{enumerate}
\vspace{-2mm} 


The rest of this article is organized as follows: \Cref{sec:motivation} provides motivation for this study. \Cref{sec:methodology} introduces our methodology for enabling mixed-precision in SEM-based CFD simulations, while \Cref{sec:tgv} presents the Taylor-Green Vortex case. \Cref{sec:results} demonstrates the results from both accuracy and performance perspectives. Finally, \Cref{sec:conclusion} discusses the main outcomes and outlines future work.

\subsection{Motivation}
\label{sec:motivation}
In the existing studies~\cite{anztsurvey,theo22}, the focus is typically placed on the algorithm itself, and the performance is often evaluated under analytically chosen stopping tolerances, such as $10^{-8}$, $10^{-10}$, or $10^{-12}$. Under these settings, the iteration count may vary substantially across problem instances, ranging, for example, from only a few iterations to several hundred, depending on the conditioning of the problem and the quality of the preconditioner. Such scenarios naturally provide room for fine-grained mixed-precision optimization, because the reduced-precision components are repeatedly exercised inside relatively long Krylov iterations and can therefore accumulate meaningful savings. This was also our observation on solving the Poisson's equation with Nekbone~\cite{ppam24} and Neko~\cite{fgcs26}, where we employed a computer arithmetic tool Verificarlo~\cite{Denis2016verificarlo} to examine actual precision requirements with the preconditioned CG during the entire run.

However, this picture changes in some spectral element method (SEM)-based CFD simulations, where the overall application involves a few coupled Krylov solvers rather than a single dominant linear solve. For instance, in the Taylor–Green Vortex (TGV) case, described in detail in~\Cref{sec:tgv}, the velocity components in the X-, Y-, and Z-directions are solved using CG with the Jacobi preconditioner, typically with an application-standard tolerance of $10^{-6}$ or $10^{-7}$. In contrast, the pressure system is solved using GMRES preconditioned by a three-level hybrid Schwarz multigrid (HSMG) method; moreover, within HSMG, the coarse-grid problem is itself solved by CG with the Jacobi preconditioner. The pressure solve uses a significantly looser tolerance, typically $10^{-4}$ or $10^{-5}$.

\vspace{-1em}
\setlength{\heavyrulewidth}{1.0pt}
\setlength{\lightrulewidth}{0.5pt}
\setlength{\arrayrulewidth}{0.5pt}
\setlength{\aboverulesep}{0pt}
\setlength{\belowrulesep}{0pt}
\begin{table}[!htbp]
\centering
\caption{Statistics of TGV tests ($Re=360$) including the number of iterations for two solvers, fluid step time, and the number of such steps. TGV-x corresponds to TGV with the polynomial degree x.}
\label{tab:tgv-p-refinement}
\begin{tabular}{ccccc}
\toprule
\multirow{2}{*}{Test Case} & \multicolumn{2}{c}{Iterations (min/max/avg/total)} & \multirow{2}{*}{\makecell{Fluid Step Time (s) \\ (min/max/avg/total)}} & \multirow{2}{*}{Steps} \\
\cline{2-3}
 & Velocity & Pressure &  &  \\
\midrule
TGV-5 & 2/3/3/5917 & 1/5/1/2093 & 0.02/0.28/0.03/59.70 & 2037 \\
TGV-6 & 3/3/3/8457 & 1/5/1/2859 & 0.03/0.35/0.04/112.83 & 2819 \\
TGV-7 & 3/3/3/11193 & 1/5/1/3764 & 0.04/0.55/0.06/205.78 & 3731\\
TGV-8 & 3/3/3/14319 & 1/5/1/4806 & 0.06/3.58/0.07/356.82 & 4773 \\
\bottomrule
\end{tabular}
\end{table}
\vspace{-1em}

\begin{figure}[!ht]
\centering
\resizebox{\textwidth}{!}{
\begin{tikzpicture}
\begin{groupplot}[
    group style={
        group size=1 by 2,
        vertical sep=1.0cm
    },
    width=\columnwidth,
    height=0.30\columnwidth,
    xmin=1, xmax=3726, 
    enlarge x limits={abs=30},
    tick align=outside,
    scaled ticks=false,
    every axis plot/.append style={
        no markers,
    },
    tick label style={font=\footnotesize},
    label style={font=\footnotesize},
    legend style={
        font=\footnotesize,
        draw=none,
        fill=none,
        at={(0.5,1.15)},
        anchor=south,
        legend columns=2,
        /tikz/every even column/.append style={column sep=0.5cm},
    },
    major grid style={line width=0.15pt, draw=gray!20},
    minor tick num=0,
    disabledatascaling,
    unbounded coords=discard,
]

\nextgroupplot[
    ylabel={Iterations},
    ymin=0.5, ymax=5.2,
    ytick={1,2,3,4,5},
    grid=major,
]

\addplot[
    thick,
    color=siamblue,
    const plot,
    mark=*,
    mark size=1.4pt,
] table[
    col sep=comma,
    x=step,
    y=pressure_iters
]{files/neko-gpu-tgv-dp-p7.csv};
\addlegendentry{Pressure Solver}

\addplot[
    thick,
    dashed,
    color=siamorange,
    mark=*,
    mark size=1.4pt,
] table[
    col sep=comma,
    x=step,
    y=x-velocity_iters
]{files/neko-gpu-tgv-dp-p7.csv};
\addlegendentry{Velocity Solver}

\nextgroupplot[
    xlabel={Step},
    ylabel={Fluid-step time (s)},
    ymode=log,
    log basis y={10},
    grid=major,
]

\addplot[
    thick,
    black!90,
] table[
    col sep=comma,
    x=step,
    y=total_step_time 
]{files/sampled_neko-gpu-tgv-dp-p7.csv};

\end{groupplot}
\end{tikzpicture}
}
\caption{Pressure and \(x\)-velocity iteration counts (top) and fluid-step time (bottom) across simulation steps for the TGV-7 case.}
\label{fig:solver_iters_time}
\end{figure}
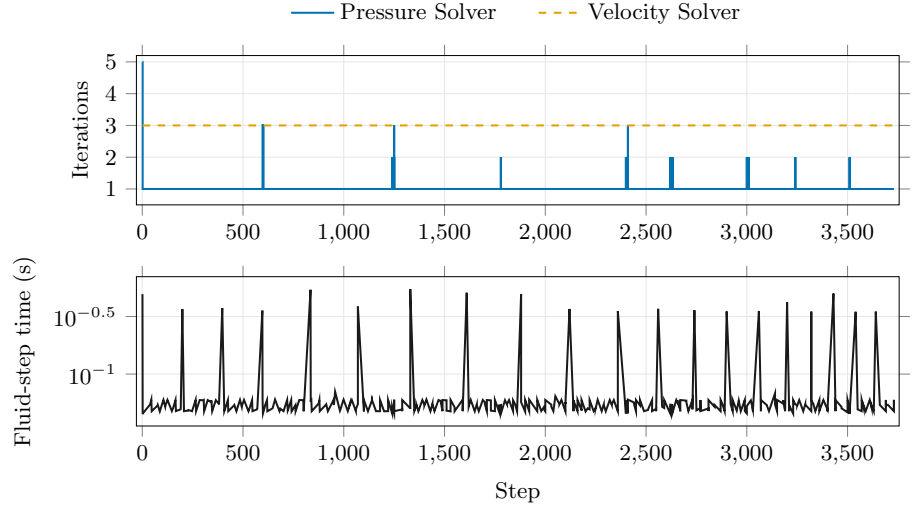

From the perspective of the full simulation, the velocity solve dominate the Krylov workload. As shown in~\Cref{tab:tgv-p-refinement} line 3, the cumulative number of CG iterations is 11,193, compared against the 3,764 total GMRES iterations for the pressure solver. A more detailed view on the number of solvers iterations per each time step and the aggregated total time step is depicted in~\Cref{fig:solver_iters_time} for the TGV case with the commonly used polynomial degree 7. This imbalance is also reflected in the profiling results, see~\Cref{fig:neko-workflow}: the velocity solve accounts for about $19\%$ of the total simulation time, whereas the pressure solve contributes only about $7\%$. In addition, the pressure solve is relatively tolerant to changes in the solver tolerance, while the velocity solve is noticeably more sensitive. Taken together, these observations indicate that TGV is a CG-dominant case. \Cref{fig:cg-gmres-residual-history} shows that the velocity and pressure solves require only a few Krylov iterations in the TGV-7 case.

\pgfplotsset{
    step1/.style   ={black, solid,  mark=o,        mark size=2.2pt, mark options={fill=white}, line width=0.75pt},
    step2/.style   ={black, dashed, mark=o,        mark size=2.2pt, mark options={fill=white}, line width=0.75pt},
    step3/.style   ={black, dotted, mark=o,        mark size=2.2pt, mark options={fill=white}, line width=0.75pt},
    step1449/.style={black, solid,  mark=square,   mark size=2.2pt, mark options={fill=white}, line width=0.75pt},
    step1450/.style={black, dashed, mark=square,   mark size=2.2pt, mark options={fill=white}, line width=0.75pt},
    step1451/.style={black, dotted, mark=square,   mark size=2.2pt, mark options={fill=white}, line width=0.75pt},
    step3729/.style={black, solid,  mark=triangle, mark size=2.4pt, mark options={fill=white}, line width=0.75pt},
    step3730/.style={black, dashed, mark=triangle, mark size=2.4pt, mark options={fill=white}, line width=0.75pt},
    step3731/.style={black, dotted, mark=triangle, mark size=2.4pt, mark options={fill=white}, line width=0.75pt},
}

\begin{figure}[!ht]
\centering

\begin{tikzpicture}
\begin{axis}[
    hide axis,
    width=0pt,
    height=0pt,
    scale only axis,
    xmin=0, xmax=1,
    ymin=0, ymax=1,
    legend columns=1,
    legend to name=sharedlegend,
    legend style={
        draw=none,
        fill=none,
        font=\scriptsize,
        row sep=1pt,
    },
    legend cell align=left,
]
\addlegendimage{step1}    \addlegendentry{Step 1}
\addlegendimage{step2}    \addlegendentry{Step 2}
\addlegendimage{step3}    \addlegendentry{Step 3}
\addlegendimage{step1449} \addlegendentry{Step 1449}
\addlegendimage{step1450} \addlegendentry{Step 1450}
\addlegendimage{step1451} \addlegendentry{Step 1451}
\addlegendimage{step3729} \addlegendentry{Step 3729}
\addlegendimage{step3730} \addlegendentry{Step 3730}
\addlegendimage{step3731} \addlegendentry{Step 3731}
\end{axis}
\end{tikzpicture}

\resizebox{\textwidth}{!}{%
\begin{tabular}{@{}c@{\hspace{0.025\textwidth}}c@{\hspace{0.018\textwidth}}c@{}}

\begin{tikzpicture}
\begin{semilogyaxis}[
    title={CG},
    title style={font=\small},
    width=0.32\textwidth,
    height=0.285\textwidth,
    scale only axis,
    xmin=0.8, xmax=3.2,
    xtick={1,2,3},
    xlabel={Iteration},
    ylabel={Residual},
    ymin=1e-7, ymax=5e-5,
    ytick={1e-7,1e-6,1e-5},
    minor tick num=0,
    minor y tick num=0,
    yminorgrids=false,
    grid=major,
    major grid style={gray!35,line width=0.3pt},
    axis line style={black,line width=0.4pt},
    tick align=inside,
    tick style={black,line width=0.4pt},
    tick label style={font=\footnotesize},
    label style={font=\small},
]

\addplot+[step1]
table[x index=0,y index=1,comment chars={\#}]
{residual_data/residual_cg_step000001.dat};

\addplot+[step2]
table[x index=0,y index=1,comment chars={\#}]
{residual_data/residual_cg_step000002.dat};

\addplot+[step3]
table[x index=0,y index=1,comment chars={\#}]
{residual_data/residual_cg_step000003.dat};

\addplot+[step1449]
table[x index=0,y index=1,comment chars={\#}]
{residual_data/residual_cg_step001449.dat};

\addplot+[step1450]
table[x index=0,y index=1,comment chars={\#}]
{residual_data/residual_cg_step001450.dat};

\addplot+[step1451]
table[x index=0,y index=1,comment chars={\#}]
{residual_data/residual_cg_step001451.dat};

\addplot+[step3729]
table[x index=0,y index=1,comment chars={\#}]
{residual_data/residual_cg_step003729.dat};

\addplot+[step3730]
table[x index=0,y index=1,comment chars={\#}]
{residual_data/residual_cg_step003730.dat};

\addplot+[step3731]
table[x index=0,y index=1,comment chars={\#}]
{residual_data/residual_cg_step003731.dat};

\end{semilogyaxis}
\end{tikzpicture}
&

\begin{tikzpicture}
\begin{semilogyaxis}[
    title={GMRES},
    title style={font=\small},
    width=0.32\textwidth,
    height=0.285\textwidth,
    scale only axis,
    xmin=0.8, xmax=5.2,
    xtick={1,2,3,4,5},
    xlabel={Iteration},
    ymin=5e-9, ymax=5e-3,
    ytick={1e-8,1e-7,1e-6,1e-5,1e-4,1e-3},
    minor tick num=0,
    minor x tick num=0,
    minor y tick num=0,
    minor tick style={draw=none},
    yminorgrids=false,
    grid=major,
    major grid style={gray!35,line width=0.3pt},
    axis line style={black,line width=0.4pt},
    tick align=inside,
    tick style={black,line width=0.4pt},
    tick label style={font=\footnotesize},
    label style={font=\small},
]

\addplot+[step1]
table[x index=0,y index=1,comment chars={\#}]
{residual_data/residual_gmres_step000001.dat};

\addplot+[step2]
table[x index=0,y index=1,comment chars={\#}]
{residual_data/residual_gmres_step000002.dat};

\addplot+[step3]
table[x index=0,y index=1,comment chars={\#}]
{residual_data/residual_gmres_step000003.dat};

\addplot+[step1449]
table[x index=0,y index=1,comment chars={\#}]
{residual_data/residual_gmres_step001449.dat};

\addplot+[step1450]
table[x index=0,y index=1,comment chars={\#}]
{residual_data/residual_gmres_step001450.dat};

\addplot+[step1451]
table[x index=0,y index=1,comment chars={\#}]
{residual_data/residual_gmres_step001451.dat};

\addplot+[step3729]
table[x index=0,y index=1,comment chars={\#}]
{residual_data/residual_gmres_step003729.dat};

\addplot+[step3730]
table[x index=0,y index=1,comment chars={\#}]
{residual_data/residual_gmres_step003730.dat};

\addplot+[step3731]
table[x index=0,y index=1,comment chars={\#}]
{residual_data/residual_gmres_step003731.dat};

\end{semilogyaxis}
\end{tikzpicture}
&

\pgfplotslegendfromname{sharedlegend}

\end{tabular}%
}

\vspace{0.6em}

\caption{Residual histories for selected fluid steps of the TGV-7 case ($Re=360$). Left: CG residuals in the x-velocity solver, averaging approximately three iterations. Right: GMRES residuals in the pressure solver, with five iterations at the first selected step and one iteration for the subsequent selected steps.}
\label{fig:cg-gmres-residual-history}
\end{figure}
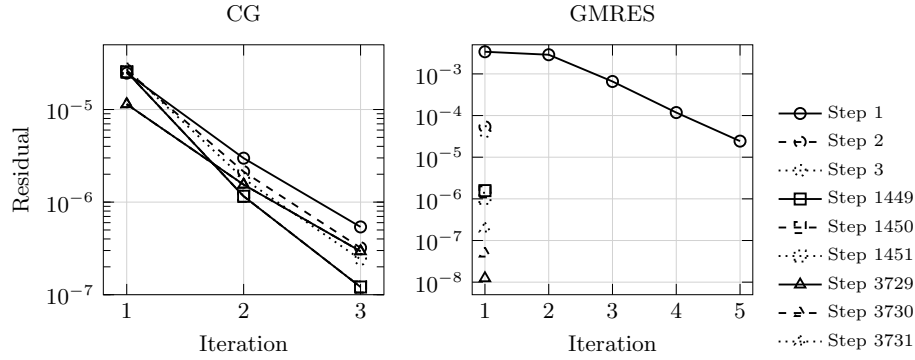

\begin{figure}[!ht]
\resizebox{\textwidth}{!}{
\begin{tikzpicture}[
    font=\large,
    node distance=4.5mm and 8mm,
    box/.style={
        draw,
        rounded corners,
        align=center,
        minimum width=8.8cm,
        minimum height=1.0cm,
        inner sep=3pt
    },
    smallbox/.style={
        draw,
        rounded corners,
        align=center,
        minimum width=3.9cm,
        minimum height=1.55cm,
        inner sep=3pt
    },
    stage/.style={box, fill=blue!5},
    sem/.style={box, fill=red!6},
    krylov/.style={smallbox, fill=green!8},
    prec/.style={smallbox, fill=yellow!12},
    state/.style={box, fill=orange!10},
    semgroup/.style={
        draw,
        dotted,
        rounded corners,
        inner sep=5pt
    },
    group/.style={
        draw,
        dashed,
        rounded corners,
        inner sep=5pt
    },
    line/.style={->, thick},
    leftnote/.style={
        font=\large\bfseries,
        align=right,
        text width=4.5cm
    },
    rightnote/.style={
        font=\large\bfseries,
        align=left
    }
]

\node[stage] (pre) {\textbf{Preprocessing / initialization}\\
mesh, geometry, SEM operators, initial condition, solver setup};

\node[stage, below=6mm of pre] (loop) {\textbf{Time-stepping loop}\\
for $n=0,1,\dots,N_t-1$};

\node[sem, below=6mm of loop] (rhs) {\textbf{SEM level: explicit RHS evaluation}\\
dealiased advection: 33.8\% \quad RHS maker: 2.7\%\\
BC: 1.3\%};

\node[krylov, below left=10mm and 2mm of rhs] (cg) {\textbf{CG solver}\\
18.3\%};

\node[prec, right=10mm of cg] (jac) {\textbf{Jacobi preconditioner}\\
1.9\%};

\node[group, fit=(cg)(jac)] (velgroup) {};

\node[leftnote, anchor=east] at ([xshift=-6mm]velgroup.west)
{Velocity solve: solver + preconditioner};

\node[state, below=12mm of velgroup] (ustar) {\textbf{Intermediate velocity} $\mathbf{u}^*$};

\node[sem, below=6mm of ustar] (divu) {\textbf{SEM level: pressure setup / residual build}\\
projection pre-solving: 4.9\% \quad pressure residual compute: 14.0\%};

\node[krylov, below left=10mm and 2mm of divu] (gmres) {\textbf{GMRES solver}\\
7.7\%};

\node[prec, right=10mm of gmres] (hsmg) {\textbf{HSMG preconditioner}\\
5.7\%};

\node[group, fit=(gmres)(hsmg)] (prsgroup) {};

\node[leftnote, anchor=east] at ([xshift=-6mm]prsgroup.west)
{Pressure solve: solver + preconditioner};

\node[sem, below=12mm of prsgroup] (proj) {\textbf{SEM level: projection / correction}\\
5.9\%};

\node[state, below=6mm of proj] (unew) {\textbf{Updated state} $\mathbf{u}^{n+1}$};

\node[stage, below=7mm of unew] (post) {\textbf{Postprocessing / output}\\
diagnostics, checkpoints, field output, visualization data};

\node[semgroup, fit=(rhs)(divu)(proj)(prsgroup)] (semops) {};

\node[leftnote, anchor=east] at ([xshift=-8mm,yshift=2mm]semops.west |- rhs.north)
{SEM-level timestep operations};

\draw[line] (pre) -- (loop);
\draw[line] (loop) -- (rhs);

\draw[line] (rhs) -- (velgroup.north);
\draw[line] (velgroup.south) -- (ustar);

\draw[line] (ustar) -- (divu);

\draw[line] (divu) -- (prsgroup.north);
\draw[line] (prsgroup.south) -- (proj);

\draw[line] (proj) -- (unew);
\draw[line] (unew) -- (post);

\coordinate (retA) at ([xshift=29mm]rhs.south |- unew.east);
\coordinate (retB) at ([xshift=29mm]rhs.south);

\draw[thick] (unew.east) -- (retA);

\draw[line] (retA) -- (retB);

\node[rightnote, anchor=west] at ([xshift=-10mm,yshift=-71mm]$(retA)!0.5!(retB)$)
{\textbf{next}\\\textbf{timestep}};

\end{tikzpicture}
}
\caption{Workflow of the SEM-based Taylor--Green vortex (TGV) simulation in Neko, annotated with CPU-time hotspot percentages from profiling with Intel VTune. Preprocessing and postprocessing surround a repeated time-stepping simulation loop. Within each timestep, SEM-type operations assemble the explicit RHS, build the pressure residual, and apply the projection update. The velocity solve is represented by the CG solver with the Jacobi preconditioner, while the pressure solve is by GMRES with the HSMG preconditioner.}
\label{fig:neko-workflow}
\end{figure}
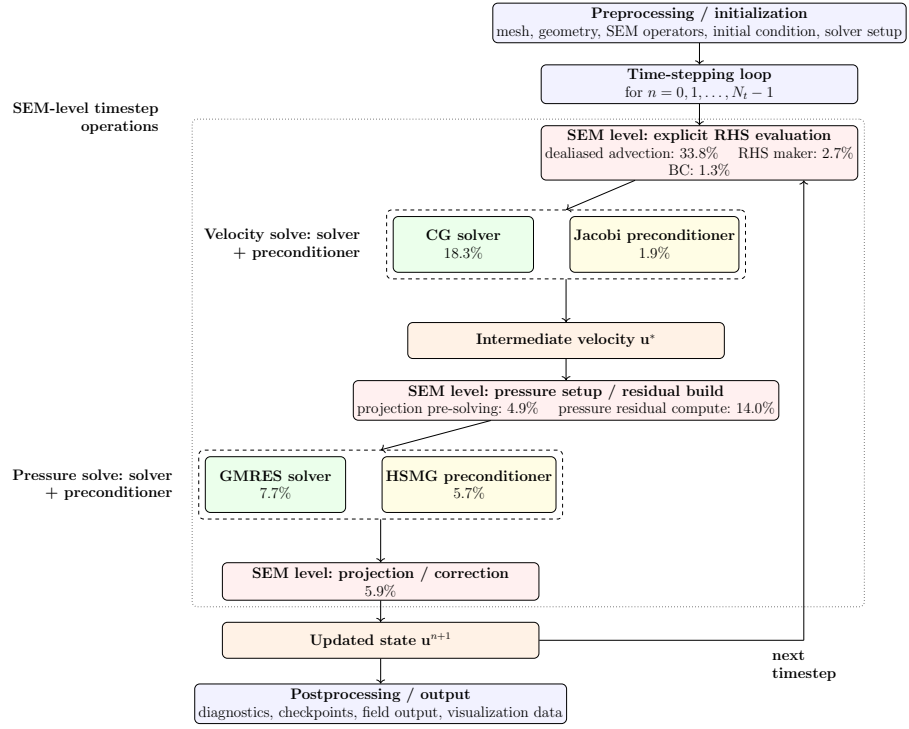

At the same time, the temporal budget available for solver-level optimization is limited. In each fluid solve time step, the average step time is only about 0.06 seconds (s), with a minimum of 0.04\,s and a maximum of 0.55\,s. This short per-step runtime and a small number of iterations per time step leave comparatively little room for fine-grained mixed-precision techniques to deliver substantial end-to-end gains, especially when those techniques target only the inner operations of individual Krylov iterations. Therefore, although fine-grained mixed-precision algorithms can be highly effective in isolated linear solver studies, their practical impact in SEM-based CFD applications may be constrained by the solver coupling, application-specific tolerances, and the limited optimization space available within each time step.

\section{Methodology}
\label{sec:methodology}
Our methodology for enabling mixed-precision in SEM-based CFD codes, e.g. Neko~\cite{neko}, NekRS~\cite{nekrs}, Nekbone~\cite{nekbone}, is based on three hierarchical levels, see~\Cref{fig:three-level-mxp-model}: block, kernel, and time-step. The upper precision limit, $\mathrm{rp}$, is imposed globally. Consequently, the precision at each of the three levels must satisfy $\operatorname{prec} \leq \mathrm{rp}$. The only exception is parallel reduction via {\tt MPI\_Allreduce} where higher precision can be used for reduction with the cap at fp64.

\begin{figure}[!ht]
\centering
\resizebox{\textwidth}{!}{
\begin{tikzpicture}[
    >=Stealth,
    line/.style={draw=black!75, line width=0.55pt},
    arrow/.style={-Stealth, draw=black!75, line width=0.6pt},
    capbox/.style={
        line,
        rounded corners=2pt,
        fill=black!3,
        minimum width=75mm,
        minimum height=19mm,
        align=center
    },
    blockbox/.style={
        line,
        rounded corners=2pt,
        fill=black!2,
        minimum width=56mm,
        minimum height=23mm,
        align=center
    },
    rolebox/.style={
        line,
        rounded corners=1.2pt,
        fill=white,
        text width=16.5mm,
        minimum height=6.1mm,
        inner sep=1pt,
        text height=1.7ex,
        text depth=.25ex,
        font=\small,
        align=center
    },
    kernelbox/.style={
        line,
        rounded corners=2pt,
        fill=black!2,
        minimum width=36mm,
        minimum height=14mm,
        align=center
    },
    stepbox/.style={
        line,
        rounded corners=2pt,
        fill=black!3,
        minimum width=115mm,
        minimum height=22mm,
        align=center
    },
    labeltext/.style={
        font=\sffamily\bfseries\large,
        align=right,
        text=black!85
    },
    note/.style={
        font=\sffamily\itshape\normalsize,
        align=left,
        text=black!80
    }
]

\node[capbox] (global) {
    {\bfseries\large Global precision cap}\\[1.1mm]
    {\large $rp \in \{\mathrm{fp64}, \mathrm{fp32}\}$}\\[1.1mm]
    {\sffamily\itshape\normalsize all selected precisions satisfy $\mathrm{prec} \leq \mathrm{rp}$}
};

\node[blockbox, below=11mm of global] (vel) {};
\node[blockbox] (sem) at ($(vel.center)+(-59mm,0)$) {};
\node[blockbox] (prs) at ($(vel.center)+(59mm,0)$) {};

\node[font=\bfseries\large] at ($(sem.center)+(0,7.5mm)$) {SEM};
\node[font=\bfseries\large] at ($(vel.center)+(0,7.5mm)$) {Velocity};
\node[font=\bfseries\large] at ($(prs.center)+(0,7.5mm)$) {Pressure};

\node[rolebox] at ($(sem.center)+(-18mm,1.0mm)$) {compute};
\node[rolebox] at ($(sem.center)+(0mm,1.0mm)$) {accumulate};
\node[rolebox] at ($(sem.center)+(18mm,1.0mm)$) {storage};

\draw[dashed, black!55, line width=0.45pt]
    ($(sem.west)+(2mm,-4.8mm)$) -- ($(sem.east)+(-2mm,-4.8mm)$);

\node[font=\sffamily\itshape\normalsize] at ($(sem.center)+(0,-8.5mm)$)
    {default block policy};

\node[rolebox] at ($(vel.center)+(-18mm,1.0mm)$) {compute};
\node[rolebox] at ($(vel.center)+(0mm,1.0mm)$) {accumulate};
\node[rolebox] at ($(vel.center)+(18mm,1.0mm)$) {storage};

\draw[dashed, black!55, line width=0.45pt]
    ($(vel.west)+(2mm,-4.8mm)$) -- ($(vel.east)+(-2mm,-4.8mm)$);

\node[font=\sffamily\itshape\normalsize] at ($(vel.center)+(0,-8.5mm)$)
    {default block policy};

\node[rolebox] at ($(prs.center)+(-18mm,1.0mm)$) {compute};
\node[rolebox] at ($(prs.center)+(0mm,1.0mm)$) {accumulate};
\node[rolebox] at ($(prs.center)+(18mm,1.0mm)$) {storage};

\draw[dashed, black!55, line width=0.45pt]
    ($(prs.west)+(2mm,-4.8mm)$) -- ($(prs.east)+(-2mm,-4.8mm)$);

\node[font=\sffamily\itshape\normalsize] at ($(prs.center)+(0,-8.5mm)$)
    {default block policy};

\node[kernelbox, below=8mm of sem] (ksem) {
    \bfseries\large hotspot kernels\\[1mm]
    \ttfamily\large tnsr3d / \dots
};

\node[kernelbox, below=8mm of vel] (kvel) {
    \bfseries\large hotspot kernels\\[1mm]
    \ttfamily\large glsc3 / \dots
};

\node[kernelbox, below=8mm of prs] (kprs) {
    \bfseries\large hotspot kernels\\[1mm]
    \ttfamily\large hsmg / \dots
};

\node[note, right=4mm of kprs] {
    kernel-level rules\\
    override block defaults
};

\node[stepbox, below=13mm of kvel] (step) {};

\node[font=\bfseries\large] at ($(step.center)+(0,6.5mm)$)
    {Step-based control};

\coordinate (barL) at ($(step.center)+(-41mm,-0.2mm)$);
\coordinate (barR) at ($(step.center)+(41mm,-0.2mm)$);

\draw[line, fill=white]
    ($(barL)+(0,-2.8mm)$) rectangle ($(barR)+(0,2.8mm)$);

\draw[line]
    ($(barL)+(27mm,-2.8mm)$) --
    ($(barL)+(31mm,0)$) --
    ($(barL)+(27mm,2.8mm)$);

\draw[line]
    ($(barL)+(55mm,-2.8mm)$) --
    ($(barL)+(59mm,0)$) --
    ($(barL)+(55mm,2.8mm)$);

\node[font=\normalsize] at ($(barL)+(13.5mm,0)$) {steps 1--100};
\node[font=\normalsize] at ($(barL)+(43mm,0)$) {steps 101--500};
\node[font=\normalsize] at ($(barL)+(70mm,0)$) {steps 501--$N$};

\node[font=\sffamily\itshape\normalsize] at ($(step.center)+(0,-7mm)$)
    {precision policy may vary over simulation steps};

\draw[arrow] (global.south) -- ++(0,-3.5mm) coordinate (gbranch);

\draw[line] (gbranch) -| (sem.north);
\draw[line] (gbranch) -- (vel.north);
\draw[line] (gbranch) -| (prs.north);

\draw[arrow] ($(sem.north)+(0,3.5mm)$) -- (sem.north);
\draw[arrow] ($(vel.north)+(0,3.5mm)$) -- (vel.north);
\draw[arrow] ($(prs.north)+(0,3.5mm)$) -- (prs.north);

\draw[arrow] (sem.south) -- (ksem.north);
\draw[arrow] (vel.south) -- (kvel.north);
\draw[arrow] (prs.south) -- (kprs.north);

\coordinate (join) at ($(kvel.south)+(0,-5mm)$);

\draw[line] (ksem.south) |- (join);
\draw[line] (kprs.south) |- (join);
\draw[arrow] (kvel.south) -- (join) -- (step.north);

\coordinate (labelx) at ($(sem.west)+(-21mm,0)$);

\node[labeltext] at (labelx |- global) {Global\\bound};
\node[labeltext] at (labelx |- sem) {Block\\level};
\node[labeltext] at (labelx |- ksem) {Kernel\\level};
\node[labeltext] at (labelx |- step) {Time/step\\level};

\draw[black!70, line width=0.5pt]
    ($(global.north west)+(-9mm,0)$) -- ++(-2mm,0) --
    ($(global.south west)+(-11mm,0)$) -- ++(2mm,0);

\draw[black!70, line width=0.5pt]
    ($(sem.north west)+(-9mm,0)$) -- ++(-2mm,0) --
    ($(sem.south west)+(-11mm,0)$) -- ++(2mm,0);

\draw[black!70, line width=0.5pt]
    ($(ksem.north west)+(-9mm,0)$) -- ++(-2mm,0) --
    ($(ksem.south west)+(-11mm,0)$) -- ++(2mm,0);

\draw[black!70, line width=0.5pt]
    ($(step.north west)+(-9mm,0)$) -- ++(-2mm,0) --
    ($(step.south west)+(-11mm,0)$) -- ++(2mm,0);

\node[note, below=1.2mm of step] {
    Precision policy:
    (block, role, kernel, step range) $\mapsto$ precision,
    with all selected precisions bounded by $rp$.
};

\end{tikzpicture}
}
\caption{Three-level hierarchical mixed-precision control model. The runtime precision $rp$ defines the global upper bound. Each program block specifies compute, accumulation, and storage precisions, while fine-grained kernel-level rules can override block defaults for hotspot kernels.}
\label{fig:three-level-mxp-model}
\end{figure}
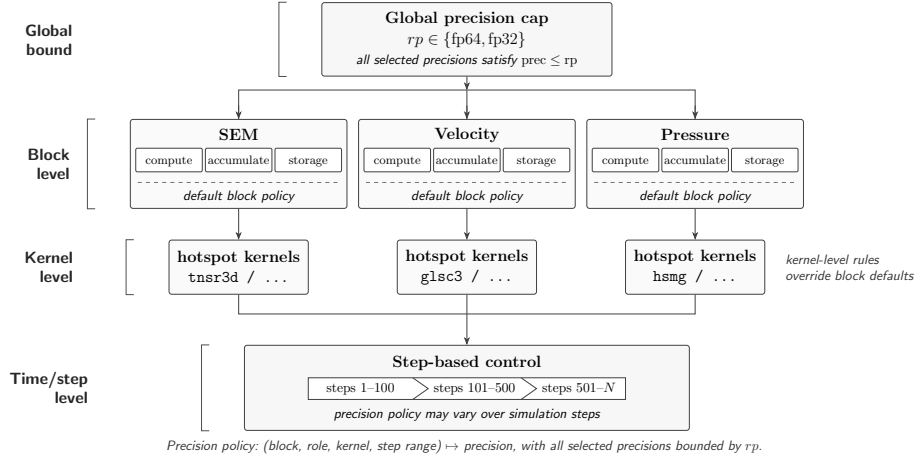

\begin{figure}[!ht]
\centering
\resizebox{\textwidth}{!}{
\begin{tikzpicture}[
    bar/.style={draw=black, line width=0.45pt, minimum height=0.55cm},
    legendbox/.style={draw=black, line width=0.45pt, minimum width=0.48cm, minimum height=0.32cm},
]

\definecolor{semcol}{RGB}{210,225,247}
\definecolor{velocitycol}{RGB}{252,242,198}
\definecolor{pressurecol}{RGB}{250,225,195}
\definecolor{otherscol}{RGB}{238,238,238}

\def\barX{3.6}        
\def\barW{10.0}       
\def\barH{0.55}       

\pgfmathsetmacro{\yA}{0}
\pgfmathsetmacro{\yB}{\yA-\barH}
\pgfmathsetmacro{\yC}{\yB-\barH}
\pgfmathsetmacro{\yD}{\yC-\barH}
\pgfmathsetmacro{\yE}{\yD-\barH}

\node[anchor=west, font=\large\sffamily] at (1.0,0.75) {Case};
\node[font=\large\sffamily] at ({\barX + 0.5*\barW},0.75) {Percentage};

\newcommand{\stackedbar}[6]{%
    \node[anchor=east] at (\barX-0.65,#2) {#1};

    \pgfmathsetmacro{\xzero}{\barX}
    \pgfmathsetmacro{\wsem}{\barW*#3/100}
    \pgfmathsetmacro{\wvel}{\barW*#4/100}
    \pgfmathsetmacro{\wpres}{\barW*#5/100}
    \pgfmathsetmacro{\woth}{\barW*#6/100}

    \pgfmathsetmacro{\xone}{\xzero+\wsem}
    \pgfmathsetmacro{\xtwo}{\xone+\wvel}
    \pgfmathsetmacro{\xthree}{\xtwo+\wpres}
    \pgfmathsetmacro{\xfour}{\xthree+\woth}

    \path[draw=black, fill=semcol, line width=0.45pt]
        (\xzero,#2-\barH/2) rectangle (\xone,#2+\barH/2);
    \node at ({(\xzero+\xone)/2},#2) {#3\%};

    \path[draw=black, fill=velocitycol, line width=0.45pt]
        (\xone,#2-\barH/2) rectangle (\xtwo,#2+\barH/2);
    \node at ({(\xone+\xtwo)/2},#2) {#4\%};

    \path[draw=black, fill=pressurecol, line width=0.45pt]
        (\xtwo,#2-\barH/2) rectangle (\xthree,#2+\barH/2);
    \node at ({(\xtwo+\xthree)/2},#2) {#5\%};

    \path[draw=black, fill=otherscol, line width=0.45pt]
        (\xthree,#2-\barH/2) rectangle (\xfour,#2+\barH/2);
    \node at ({(\xthree+\xfour)/2},#2) {#6\%};
}

\stackedbar{$N=512$, $p=6$}{\yA}{12}{19}{30}{39}
\stackedbar{$N=512$, $p=7$}{\yB}{10}{19}{32}{39}
\stackedbar{$N=512$, $p=8$}{\yC}{13}{19}{26}{42}
\stackedbar{$N=32768$, $p=6$}{\yD}{25}{21}{26}{28}
\stackedbar{$N=32768$, $p=7$}{\yE}{31}{18}{24}{27}

\def\legY{-3.05}
\def\boxW{0.52}
\def\boxH{0.3}

\path[draw=black, fill=semcol, line width=0.45pt]
    (3.6,\legY-\boxH/2) rectangle +(\boxW,\boxH);
\node[anchor=west] at (4.35,\legY) {SEM};

\path[draw=black, fill=velocitycol, line width=0.45pt]
    (5.9,\legY-\boxH/2) rectangle +(\boxW,\boxH);
\node[anchor=west] at (6.65,\legY) {Velocity};

\path[draw=black, fill=pressurecol, line width=0.45pt]
    (8.7,\legY-\boxH/2) rectangle +(\boxW,\boxH);
\node[anchor=west] at (9.45,\legY) {Pressure};

\path[draw=black, fill=otherscol, line width=0.45pt]
    (11.6,\legY-\boxH/2) rectangle +(\boxW,\boxH);
\node[anchor=west] at (12.35,\legY) {Others};

\end{tikzpicture}
}
\vspace{-1em}
\caption{Breakdown of the accumulated runtime inside each $P_N-P_N$ fluid step into SEM, velocity, pressure, and other components. 
Percentages are computed from the profiled time in the fluid solve routine over the full simulation run.}
\label{fig:case-percentage}
\end{figure}

The block level, which corresponds to the fluid solve, encapsulates SEM, velocity, and pressure parts. \Cref{fig:case-percentage} highlights the impact of these three solvers following the $P_N-P_N$ scheme~\cite{karpEffectsLowerFloatingpoint2026} on the overall runtime of Neko as a function of elements/ DoFs and a polynomial order. 
SEM includes dealiased advection, RHS assembly, boundary-condition treatment, and projection/correction; velocity includes the velocity residual and the CG+Jacobi solve; pressure includes the pressure residual and the GMRES+HSMG solve; and others collect smaller shared operations such as projection pre-solving, gather/scatter, reductions, and logging/control overhead. 
Notably, the SEM share increases 2-3 times for higher $N$.
Each part may have individual precision but preferably the identical higher precision for the cross solver synchronization within the block level. Within each part, operations on data can be divided into compute, accumulate, and storage. While accumulate is accuracy sensitive, compute and storage are more keen toward lowering precision. The latter two can also be separated -- data can be stored in different precision than the compute performs. This follows the idea similar to the memory accessor~\cite{Flegar2021Ginkgo} and further reduces the load on memory bandwidth. 

At the kernel level, we select some of the dominant kernels of the solvers defined at the block level. Here we can further explore precision tuning under the condition of overriding the block default precision setups.

Since SEM-based CFD simulation are not suitable for precision changes within the Krylov-type solvers on each individual step, we envision to change precision across different time steps. For instance, at the start of simulations when the physical system is formed, it is essential to use full precision across all three levels. However, this requirement can be lowered when the CG and GMRES settle to naturally low number of iterations per time step. 

In this article, we focus on one component of this methodology: block-level precision control, with selected exploratory studies at the kernel level for further precision reduction.

\section{Taylor-Green Vortex}
\label{sec:tgv}
The current investigation is done based on a classical benchmarking case in CFD: the so called Taylor-Green Vortex, which is commonly used to validate and verify numerical methods as done, for example, in \cite{du_gjp_2025}. 
The case considers the temporal development and decay of a three-dimensional vortex within a periodic box-like domain of size $[-L\pi,L\pi]^3$. The simulation starts from an initial velocity field defined as follows:
\begin{eqnarray}
        u(t=0) &=& v_0 \sin(x)\cos(y)\cos(z),\\
        v(t=0) &=& -v_0\cos(x)\sin(y)\cos(z),\\
        w(t=0) &=& 0.
\end{eqnarray}

In this study, three Reynolds numbers are considered: $Re=360$, $Re=1600$, and $Re=10{,}000$. The Reynolds number is defined based on the maximum magnitude of the initial velocity field $v_o$, the length scale of the largest vortical structures which is equal to $L$ and the kinematic viscosity of the fluid $\nu$:
\begin{equation}
Re = \frac{v_o L}{\nu}\text{.}
\end{equation}

The case $Re=360$ is used for profiling and background performance analysis, whereas $Re=1600$ is used as the main Taylor--Green vortex benchmark for the mixed-precision experiments. The higher-Reynolds-number case,
$Re=10{,}000$, is included as a precision-stress test. For the $Re=360$ and $Re=1600$ cases, simulations are conducted on uniform meshes with $32 \times 32 \times 32$ elements and polynomial orders ranging from $P=5$ to $P=8$.

In all TGV experiments, simulations were conducted until a physical time of 20 seconds. Time advancement is performed using adaptive time stepping with a target $CFL$ number of $0.4$, with dealiasing enabled and a fixed third-order time-integration scheme, consistent with the case-file structure used in Neko. The velocity subsystem is solved using CG with the Jacobi preconditioner and an absolute stopping tolerance of $10^{-6}$. The pressure subsystem is solved using GMRES preconditioned by HSMG, with an absolute stopping tolerance of $10^{-4}$ and a projection space dimension of $20$. In the HSMG hierarchy, the default coarse-grid solver is CG with Jacobi preconditioning. 

\begin{figure}
\centering
\resizebox{\textwidth}{!}{
\begin{tikzpicture}
\begin{groupplot}[
    group style={
        group size=1 by 2,
        vertical sep=1.3cm
    },
    width=13cm,
    height=5.8cm,
    xmin=0,
    xmax=20,
    grid=major,
    major grid style={
        draw=gray!35,
        line width=0.25pt
    },
    tick align=outside,
    tick pos=both,
    axis line style={black!75},
    tick style={black!75},
    xlabel={$t/t_c$},
    label style={font=\normalsize},
    tick label style={font=\small},
    title style={font=\normalsize},
    every axis plot/.append style={
        black,
        line width=1.1pt,
        smooth
    }
]



\nextgroupplot[
    ylabel={enst},
    title={Enstrophy},
    ymin=0,
    ymax=2.2,
    ytick={0.5,1.0,1.5,2.0},
    xlabel={$t/t_c$}
]
\addplot[
    thick
] table[
    col sep=comma,
    x=t,
    y=enst
] {files/neko-gpu-tgv-dp-p7_post.csv};

\addplot[
    only marks,
    mark=*,
    mark size=1.8pt,
    black
] coordinates {(6.43447162,2.03058781)};

\draw[densely dashed, black!65]
    (axis cs:6.43447162,0)
    -- (axis cs:6.43447162,2.03058781);

\node[
    anchor=south west,
    align=left,
    font=\small,
    fill=white,
    fill opacity=0.88,
    text opacity=1,
    rounded corners=2pt,
    inner sep=3pt
] at (axis cs:6.9,1.45)
{Peak enstrophy\\
$t/t_c = 6.43447162$\\
$\Omega_{\max}=2.03058781$};

\node[
    anchor=north east,
    align=left,
    font=\small,
    fill=white,
    fill opacity=0.85,
    text opacity=1,
    rounded corners=2pt,
    inner sep=3pt
] at (rel axis cs:0.97,0.95)
{Transition to turbulent\\
small-scale structures\\
and isotropic decay};

\end{groupplot}
\end{tikzpicture}
}
\caption{
Volume-averaged enstrophy for the Taylor--Green vortex case at $\mathrm{Re}=360$ and polynomial order $p=7$. The enstrophy reaches a peak value of $2.03058781$ at $t/t_c = 6.43447162$.
}
\label{fig:tgv-enst}
\end{figure}
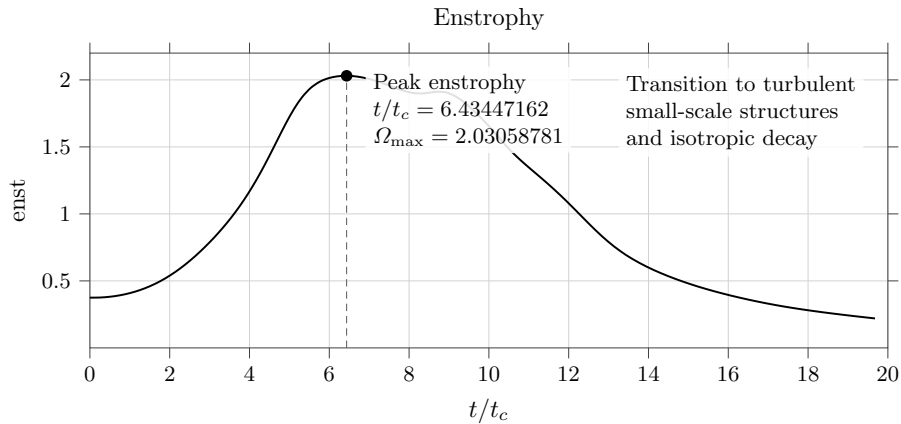

Given by the initial conditions, the flow is characterized by a smooth flow field at the beginning. In the course of time, a transition into a turbulent flow field characterized by small-scale turbulent structures and an isotropic decay takes place. In \Cref{fig:tgv-enst} the volume-averaged enstrophy is shown. While the kinetic energy undergoes a decaying process, enstrophy reaches a peak value at $t/t_c = 6.43447162$ with $t_c=L/v_0$. 


\section{Experimental results}
\label{sec:results}
In this section, we evaluate two groups of mixed-precision (MxP) configurations, comprising eight representative configurations in total. As shown in \Cref{tab:mxp-config}, A-0 and B-0 serve as the two global baseline cases, corresponding to global double precision and global single precision, respectively. The remaining configurations are designed to investigate block-level compute precision assignments for the SEM, velocity, and pressure components under different global precision caps.

All experiments were performed on the Karolina supercomputer hosted at IT4I, Czech Republic. Each node consists of two AMD Zen 3 EPYC\textsuperscript{TM} 7763 processors @2.45 GHz, 1024\,GB of system memory, and eight NVIDIA A100 GPUs. The CUDA backend code was compiled with \texttt{-arch=sm\_80 -O3}, while the Fortran components were compiled with \texttt{-O3}. The software environment used is composed of  CUDA 12.4.0, OpenMPI 4.1.6 with CUDA-aware MPI enabled, and GFortran 12.2.0. This hardware platform also supports energy measurement, which is well suited for assessing the performance implications of mixed-precision strategies in large-scale GPU-accelerated CFD simulations.

The experimental study is based on the Taylor–Green vortex benchmark, described in~\Cref{sec:tgv}, using Reynolds number of $Re=1600$ and polynomial degree of $P=8$. The mesh resolution of $32\times32\times32$ is used, corresponding to 32,768 elements. Throughout the experiments, the number of devices is chosen such that the workload per device remains within approximately 7,000 to 10,000 elements, which is consistent with common practical scaling guidelines for this class of simulations. For the $32\times32\times32$ mesh, the experiments were conducted on 4 GPUs, i.e., half of a compute node, corresponding to 8,192 elements per device. For performance measurements, we used exclusive node access to eliminate interference from co-scheduled jobs and to ensure consistency and reproducibility of the reported timings and energies. 

\begin{table}[!ht]
\vspace*{-6mm}
\centering
\caption{Mixed-precision (MxP) configurations.}
\label{tab:mxp-config}
\renewcommand{\arraystretch}{1.12}
\begin{tabular}{c c c c l}
\toprule
Config. &
SEM &
Velocity &
Pressure &
Comments \\
\midrule

\rowcolor{groupgray}
\multicolumn{5}{l}{\textbf{Group A: \texttt{rp} = fp64}} \\
0 & \multicolumn{4}{l}{Global fp64} \\
1 & fp32 & fp32 & fp32 & All fp32 compute \\
2 & fp32 & fp64 & fp64 & SEM-only fp32 compute \\
3 & fp64 & fp32 & fp64 & Velocity-only fp32 compute\\
4 & fp64 & fp64 & fp32 & Pressure-only fp32 compute \\
\midrule
\rowcolor{groupgray}
\multicolumn{5}{l}{\textbf{Group B: \texttt{rp} = fp32}} \\
0 & \multicolumn{4}{l}{Global fp32} \\
1 & fp64 & fp64 & fp64 & All fp64 compute \\
2 & fp64 & fp32 & fp32 & SEM-only fp64 compute \\
3 & fp32 & fp64 & fp32 & Velocity-only fp64 compute \\
4 & fp32 & fp32 & fp64 & Pressure-only fp64 compute \\
5 & fp32/16 & fp32 & fp32 & \texttt{adv\_dealias} (SEM) fp16 compute\\
6 & fp32 & fp32/16 & fp32 & \texttt{ax\_helm} (Velocity) fp16 compute\\
7 & fp32 & fp32 & fp32/16 & \texttt{ax\_helm} (Pressure) fp16 compute\\
\bottomrule
\end{tabular}
\end{table}

Although the mixed-precision model (see~\Cref{fig:three-level-mxp-model}) treats storage precision as an independent component at the block level, alongside compute and accumulation precision, a fully decoupled implementation of storage precision for each simulation component would require substantial architectural restructuring in Neko. To avoid this implementation overhead, in this study we adopt a simplified experimental design based on the runtime precision parameter \texttt{rp}, which specifies the global build precision in Neko.

Accordingly, we define two groups of experiments. In Group A (rp=fp64), all globally stored quantities are maintained in double precision. In Group B (rp=fp32), the global storage format is single precision. Nevertheless, the accumulation in the fp32 group is still kept in double precision to ensure numerical stability. Within each group, we further apply local storage optimizations to selected hotspot variables in the SEM, velocity, and pressure components. For instance, under the rp=fp64 configuration, selected hotspot local variables are stored in fp32, thereby reducing memory-related overhead while preserving the global fp64 software configuration. These configurations provide a practical approximation to the hierarchical mixed-precision control model without requiring intrusive software redesign.





\subsection{Performance}
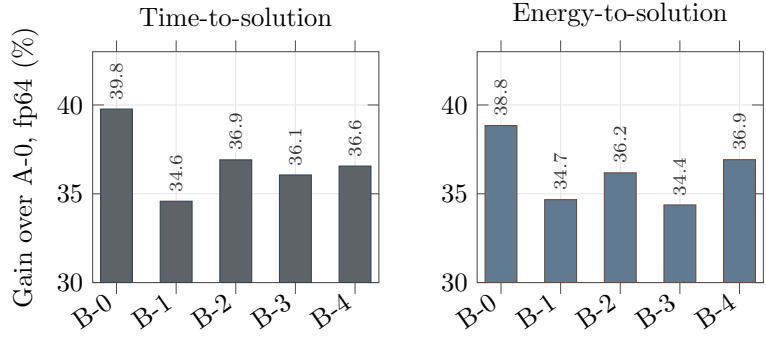
\begin{figure}[!ht]
\centering
\begin{tikzpicture}
\begin{groupplot}[
    group style={
        group size=2 by 1,
        horizontal sep=1.3cm
    },
    width=0.44\linewidth,
    height=0.38\linewidth,
    ybar,
    ymin=30,
    ymax=43,
    symbolic x coords={B-0,B-1,B-2,B-3,B-4},
    xtick=data,
    grid=major,
    major grid style={draw=gray!20},
    axis line style={black!70},
    tick style={black!70},
    x tick label style={rotate=35, anchor=east},
    nodes near coords,
    every node near coord/.append style={
        font=\scriptsize,
        rotate=90,
        anchor=west,
        black!80
    },
    point meta=explicit symbolic,
]

\nextgroupplot[
    title={Time-to-solution},
    ylabel={Gain over A-0, fp64 (\%)},
]
\addplot+[
    ybar,
    bar width=12pt,
    fill={rgb,255:red,93;green,99;blue,102},
    draw={rgb,255:red,52;green,67;blue,83}
] coordinates {
    (B-0,39.77) [39.8]
    (B-1,34.58) [34.6]
    (B-2,36.91) [36.9]
    (B-3,36.06) [36.1]
    (B-4,36.56) [36.6]
};

\nextgroupplot[
    title={Energy-to-solution},
]
\addplot+[
    ybar,
    bar width=12pt,
    fill={rgb,255:red,95;green,122;blue,145},
    draw={rgb,255:red,111;green,72;blue,55}
] coordinates {
    (B-0,38.84) [38.8]
    (B-1,34.67) [34.7]
    (B-2,36.18) [36.2]
    (B-3,34.37) [34.4]
    (B-4,36.92) [36.9]
};

\end{groupplot}
\end{tikzpicture}

\caption{Runtime and overall energy-to-solution gains for Group B configurations relative to the fp64 baseline A-0.}
\label{fig:group-b-time-energy-gain}
\end{figure}

\Cref{tab:mxp-perf-re1600-p8} reports the runtime, energy consumption, and solver iteration changes for the $Re=1600$, $P=8$ case. \Cref{fig:group-b-time-energy-gain} shows the time and energy-to-solution gain compared against fp64 baseline for Group B. Group A configurations, based on the fp64 precision bound, show similar solver counts to the fp64 baseline but slightly higher runtime and energy, indicating that the current fp32 compute paths alone do not yet overcome type casting on every time step/ solver iteration and fp64 storage overheads. In contrast, Group B configurations, based on the fp32 precision bound, substantially reduce both time- and energy-to-solution. Although the pressure iteration count increases slightly, the overall cost remains much lower, showing that fp32-based mixed precision is the more practical performance direction for this case.
\definecolor{diffplus}{RGB}{240,180,60}
\definecolor{diffminus}{RGB}{0,176,240}
\begin{table}[!htbp]
\vspace*{-6mm}
\centering
\caption{Runtime, energy consumption, and solver-count differences for the $Re=1600, P=8$ test case. Tvi and Tpi denote the total numbers of iterations for the velocity and pressure solves, respectively. Values in parentheses  indicate deviations relative to the double-precision baseline configurations A-0/ B-0.}
\label{tab:mxp-perf-re1600-p8}
\begin{tabular}{cllcccc}
\toprule
\multirow{2}{*}{Config.} & \multirow{2}{*}{Tvi (diff)} & \multirow{2}{*}{Tpi (diff)} & \multirow{2}{*}{Time (s)} & \multicolumn{3}{c}{Energy (J)} \\
\cline{5-7}
 &  &  &  & CPU & GPU & Overall \\
\midrule
A-0 & 12912 & 6777 & 596.0 & 143315 & 572017 & 715332 \\
A-1 & 12914 \textcolor{diffplus}{(+2)} & 6792 \textcolor{diffplus}{(+15)} & 579.2 & 152150 & 582132 & 734282 \\
A-2 & 12912 & 6786 \textcolor{diffplus}{(+9)} & 578.7 & 145326 & 608505 & 753831 \\
A-3 & 12910 \textcolor{diffminus}{(-2)} & 6785 \textcolor{diffplus}{(+8)} & 583.5 & 145484 & 596761 & 742245 \\
A-4 & 12912 & 6789 \textcolor{diffplus}{(+12)} & 579.5 & 144887 & 596088 & 740975 \\
\midrule
B-0 & 12910 \textcolor{diffminus}{(-2)} & 6801 \textcolor{diffplus}{(+24)} & 359.0 & 88436 & 349105 & 437541 \\
B-1 & 12910 \textcolor{diffminus}{(-2)} & 6788 \textcolor{diffplus}{(+11)} & 389.9 & 91262 & 376085 & 467347 \\
B-2 & 12912 & 6796 \textcolor{diffplus}{(+19)} & 376.0 & 95370 & 361125 & 456495 \\
B-3 & 12910 \textcolor{diffminus}{(-2)} & 6798 \textcolor{diffplus}{(+21)} & 381.1 & 97615 & 371846 & 469461 \\
B-4 & 12911 \textcolor{diffminus}{(-1)} & 6799 \textcolor{diffplus}{(+22)} & 378.1 & 90822 & 360378 & 451200 \\
\bottomrule
\end{tabular}
\end{table}

\subsection{Accuracy}
For the accuracy assessment, we select the case $Re=10{,}000$ with a polynomial order $P=7$ as a representative example. We note that this configuration is not intended to serve as a DNS-quality reference for the Taylor--Green vortex at $Re=10{,}000$. At this Reynolds number, the smallest flow scales are substantially more demanding than in the $Re=1600$ benchmark, and a DNS-quality simulation would require a substantially finer mesh than the one used here. Instead, we use this test case as a precision-stress test to evaluate whether the proposed mixed-precision configurations can remain closer to the double-precision solution than the global single-precision baseline.

A convenient way to assess the accuracy of each mixed-precision run is to compare integral quantities against a fp64 reference solution. In this work, we use the fp64 run as the baseline and evaluate the relative percentage error of the kinetic energy and enstrophy over time. These quantities are useful because they summarize the global flow evolution and are commonly used diagnostics for the Taylor-Green vortex.

Let $E^{dp}_{kin}(t)$ and $\Omega^{\mathrm{dp}}(t)$ denote the kinetic energy and enstrophy from the fp64 reference simulation. For a mixed-precision or fp32 run, the corresponding quantities are denoted by $E^{mxp}_{kin}(t)$ and $\Omega^{\mathrm{mxp}}(t)$. The relative percentage errors are computed as
\begin{equation}
e_E(t)
=
\frac{
\left| E_{\mathrm{kin}}^{\mathrm{mxp}}(t)
      - E_{\mathrm{kin}}^{\mathrm{dp}}(t) \right|
}{
\max_{t \in [0,T]}
\left| E_{\mathrm{kin}}^{\mathrm{dp}}(t) \right|
}
\times 100\%.
\label{eq:error_energy}
\end{equation}

\begin{equation}
e_\Omega(t)
=
\frac{
\left| \Omega^{\mathrm{mxp}}(t)
      - \Omega^{\mathrm{dp}}(t) \right|
}{
\max_{t \in [0,T]}
\left| \Omega^{\mathrm{dp}}(t) \right|
}
\times 100\%.
\label{eq:error_enst}
\end{equation}

Besides the time-history error curves, we also report extra three scalar error metrics to quantify the difference between the mixed-precision results and the fp64 reference. For kinetic energy, we define the maximum error as:
\begin{equation}
e_{E,\max}
=
\frac{
\max_i
\left|
E_{\mathrm{kin}}^{\mathrm{mxp}}(t_i)
-
E_{\mathrm{kin}}^{\mathrm{dp}}(t_i)
\right|
}{
\max_i
\left|
E_{\mathrm{kin}}^{\mathrm{dp}}(t_i)
\right|
}
\times 100\%.
\label{eq:e_energy_max}
\end{equation}

For enstrophy, the maximum error is defined as:
\begin{equation}
e_{\Omega,\max}
=
\frac{
\max_i
\left|
\Omega^{\mathrm{mxp}}(t_i)
-
\Omega^{\mathrm{dp}}(t_i)
\right|
}{
\max_i
\left|
\Omega^{\mathrm{dp}}(t_i)
\right|
}
\times 100\%.
\label{eq:e_enstrophy_max}
\end{equation}

Since the enstrophy peak is sensitive to the small-scale vortical structures in the Taylor--Green vortex, we also report the peak enstrophy error:
\begin{equation}
e_{\Omega,\mathrm{peak}}
=
\frac{
\left|
\max_i \Omega^{\mathrm{mxp}}(t_i)
-
\max_i \Omega^{\mathrm{dp}}(t_i)
\right|
}{
\max_i
\left|
\Omega^{\mathrm{dp}}(t_i)
\right|
}
\times 100\%.
\label{eq:e_enstrophy_peak}
\end{equation}

\begin{figure}[!ht]
\vspace*{-4mm}
\centering
\resizebox{\textwidth}{!}{
\begin{tikzpicture}
\begin{groupplot}[
    group style={
        group size=1 by 2,
        vertical sep=1.5cm,
    },
    width=13cm,
    height=5.4cm,
    grid=both,
    xmin=0,
    xmax=20,
    legend pos=north west,
    legend cell align=left,
    tick align=outside,
    tick pos=left,
    axis line style={black, line width=0.35pt},
    tick style={black, line width=0.35pt},
    major tick length=2.5pt,
    minor tick length=1.5pt,
    every axis plot/.append style={
        mark=none,
        line width=0.85pt,
    },
    major grid style={
        draw=gray!30,
        line width=0.25pt,
    },
    minor grid style={
        draw=gray!15,
        line width=0.15pt,
    },
    title style={
        font=\normalsize,
    },
    label style={
        font=\normalsize,
    },
    tick label style={
        font=\small,
    },
    legend style={
        font=\small,
        draw=black,
        line width=0.3pt,
        fill=white,
        fill opacity=0.9,
        text opacity=1,
        cells={anchor=west},
    },
]

\nextgroupplot[
    title={Group A},
    ylabel={enst error (\%)},
]

\addplot[
    color=blue!70!black,
    solid,
] table[
    col sep=comma,
    x=t,
    y=sp_enst_pct_err
] {files/neko-gpu-tgv-errors-mxp-a.csv};
\addlegendentry{fp32}

\addplot[
    color=red!75!black,
    dashed,
] table[
    col sep=comma,
    x=t,
    y=mxp_a1_enst_pct_err
] {files/neko-gpu-tgv-errors-mxp-a.csv};
\addlegendentry{mxp-a1}

\addplot[
    color=gray!45!black,
    dashdotted,
] table[
    col sep=comma,
    x=t,
    y=mxp_a2_enst_pct_err
] {files/neko-gpu-tgv-errors-mxp-a.csv};
\addlegendentry{mxp-a2}

\addplot[
    color=violet!80!black,
    densely dashed,
] table[
    col sep=comma,
    x=t,
    y=mxp_a3_enst_pct_err
] {files/neko-gpu-tgv-errors-mxp-a.csv};
\addlegendentry{mxp-a3}

\addplot[
    color=orange!85!black,
    densely dashdotted,
] table[
    col sep=comma,
    x=t,
    y=mxp_a4_enst_pct_err
] {files/neko-gpu-tgv-errors-mxp-a.csv};
\addlegendentry{mxp-a4}

\nextgroupplot[
    title={Group B},
    xlabel={$t/t_c$},
    ylabel={enst error (\%)},
]

\addplot[
    color=blue!70!black,
    solid,
] table[
    col sep=comma,
    x=t,
    y=sp_enst_pct_err
] {files/neko-gpu-tgv-errors-mxp-b.csv};
\addlegendentry{fp32}

\addplot[
    color=red!75!black,
    dashed,
] table[
    col sep=comma,
    x=t,
    y=mxp_b1_enst_pct_err
] {files/neko-gpu-tgv-errors-mxp-b.csv};
\addlegendentry{mxp-b1}

\addplot[
    color=gray!45!black,
    dashdotted,
] table[
    col sep=comma,
    x=t,
    y=mxp_b2_enst_pct_err
] {files/neko-gpu-tgv-errors-mxp-b.csv};
\addlegendentry{mxp-b2}

\addplot[
    color=violet!80!black,
    densely dashed,
] table[
    col sep=comma,
    x=t,
    y=mxp_b3_enst_pct_err
] {files/neko-gpu-tgv-errors-mxp-b.csv};
\addlegendentry{mxp-b3}

\addplot[
    color=orange!85!black,
    densely dashdotted,
] table[
    col sep=comma,
    x=t,
    y=mxp_b4_enst_pct_err
] {files/neko-gpu-tgv-errors-mxp-b.csv};
\addlegendentry{mxp-b4}

\addplot[
    color=brown!70!black,
    densely dotted,
] table[
    col sep=comma,
    x=t,
    y=mxp_b5_enst_pct_err
] {files/neko-gpu-tgv-errors-mxp-b.csv};
\addlegendentry{mxp-b5}

\addplot[
    color=cyan!60!black,
    loosely dashed,
] table[
    col sep=comma,
    x=t,
    y=mxp_b6_enst_pct_err
] {files/neko-gpu-tgv-errors-mxp-b.csv};
\addlegendentry{mxp-b6}

\addplot[
    color=magenta!70!black,
    loosely dashdotted,
] table[
    col sep=comma,
    x=t,
    y=mxp_b7_enst_pct_err
] {files/neko-gpu-tgv-errors-mxp-b.csv};
\addlegendentry{mxp-b7}

\end{groupplot}
\end{tikzpicture}
}
\caption{Enstrophy relative error compared against fp64 for Groups A and B.}
\label{fig:enstrophy-error-groups-a-b}
\vspace*{-4mm}
\end{figure}
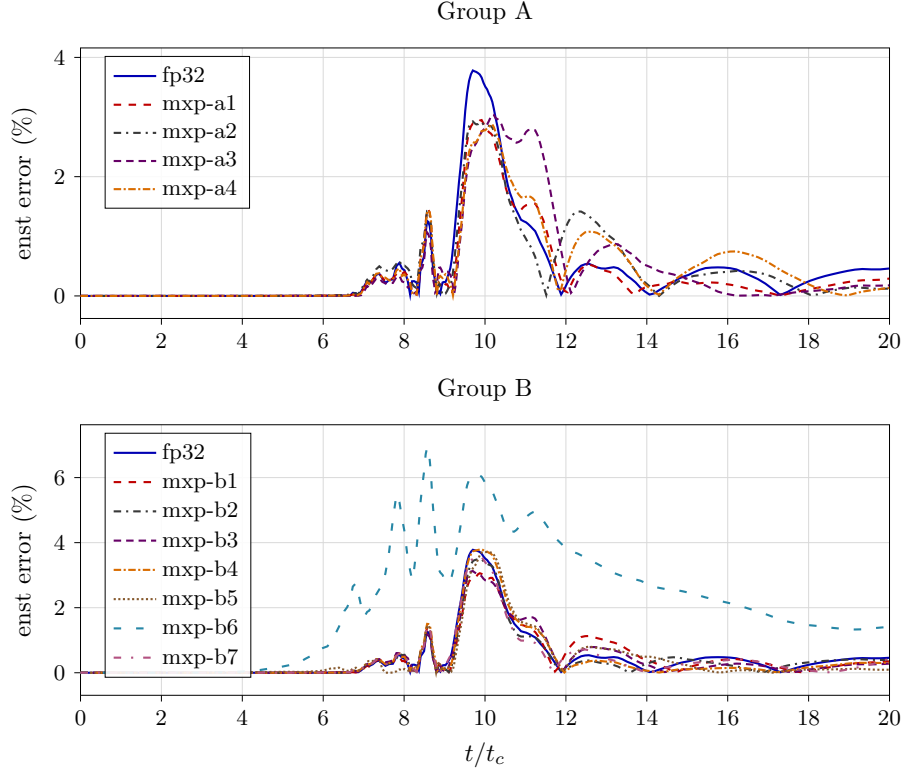

\Cref{fig:enstrophy-error-groups-a-b} show the relative error of enstrophy for Groups A and B, both includes global fp64 and fp32 as a reference. \Cref{tab:summary-metrics-max-dp} shows the summary of three metrics, $e_{E,\max}$, $e_{\Omega,\max}$, and $e_{\Omega,\mathrm{peak}}$. Group A mainly improves or preserves accuracy relative to the lower-precision alternatives. Among these configurations, A-2 (mxp-a2) gives the smallest kinetic-energy maximum error with $e_{E,\max}=0.5462\%$, while A-4 (mxp-a4) gives the smallest enstrophy maximum error, with $e_{\Omega,\max}=2.8763\%$. For the enstrophy peak-time error, A-3 (mxp-a3) gives the smallest value, $e_{\Omega,peak}=0.0104\%$. These results indicate that the most accuracy-preserving choice depends on the diagnostic quantity. The SEM-focused fp32 compute strategy, A-2, remains a promising candidate because it gives the best kinetic-energy accuracy and competitive enstrophy accuracy. Although A-2 does not yet provide a time or energy benefit in the current implementation, it is an important target for future optimization, especially once lower-precision storage becomes truly resident throughout the whole GPU time-stepping loop. In that setting, A-2 could offer a better balance between accuracy and performance by reducing SEM cost without introducing precision loss.

Group B, on the other hand, is primarily associated with performance. The B configurations have significantly lower time- and energy-to-solution than Group A, reflecting the benefit of global fp32 storage. The all-fp64 compute configuration B-1 (mxp-b1) gives the lowest enstrophy maximum error, $e_{\Omega,\max}=3.0603\%$, while the velocity-focused fp64 configuration B-3 (mxp-b3) gives the lowest enstrophy peak-time error, $e_{\Omega,peak}=0.1174\%$. This suggests that selectively increasing the compute precision in key parts of the algorithm, such as the SEM, velocity, and pressure blocks, or even the velocity block alone, can improve robustness compared with pure single precision. Notably, B-6 (mxp-b6), where fp16 is applied to the velocity-side \texttt{ax\_helm} kernel, achieves a low $e_{E,\max}$, although it produces relatively larger $e_{\Omega,\max}$ and $e_{\Omega,peak}$ errors. This indicates that fp16 may still be useful for selected kernels in CFD simulations, but its impact is strongly quantity-dependent and should be evaluated carefully, especially for gradient-sensitive diagnostics such as enstrophy.

Overall, these results suggest a two-direction mixed-precision strategy. Group A is useful for identifying accuracy-sensitive components and for developing future fp32-storage optimizations under an \texttt{rp=fp64} environment. Group B is more attractive for practical performance, especially when global single precision is fast but may become less reliable for more sensitive cases. Although B-1 and B-3 require more time and energy than the pure fp32 baseline, they provide improved numerical robustness and may be valuable for high-Reynolds-number simulations where global single precision becomes increasingly sensitive. The fp16 result in B-6 is particularly interesting for kinetic energy, but its larger enstrophy error indicates that fp16 should remain a targeted kernel-level option rather than a general replacement for fp32. Future work will therefore focus on cases with much larger single-precision error ($5-10\%$ or higher, this is a common upper bound of industry level), where the benefit of selectively restoring higher precision is expected to be more visible.

\begin{table}[!ht]
\centering
\caption{Summary of accuracy metrics for the mixed-precision configurations.}
\label{tab:summary-metrics-max-dp}
\vspace{-0.5em}
\begin{subtable}[t]{0.48\textwidth}
\centering
\caption{MxP Group A}
\label{tab:summary-metrics-group-a}
\resizebox{\textwidth}{!}{%
\begin{tabular}{lccc}
\toprule
Config & $e_{E,\max}$ (\%) & $e_{\Omega,\max}$ (\%) & $e_{\Omega,\mathrm{peak}}$ (\%) \\
\midrule
fp32   & 0.7198 & 3.7806 & 0.1764 \\
mxp-a1 & 0.6634 & 2.9493 & 0.2479 \\
mxp-a2 & \cellcolor{gray!20}\textbf{0.5462} & 2.9270 & 0.2769 \\
mxp-a3 & 0.8079 & 3.0438 & \cellcolor{gray!20}\textbf{0.0104} \\
mxp-a4 & 0.6362 & \cellcolor{gray!20}\textbf{2.8763} & 0.3166 \\
\bottomrule
\end{tabular}%
}
\end{subtable}
\hfill
\begin{subtable}[t]{0.48\textwidth}
\centering
\caption{MxP Group B}
\label{tab:summary-metrics-group-b}
\resizebox{\textwidth}{!}{%
\begin{tabular}{lccc}
\toprule
Config & $e_{E,\max}$ (\%) & $e_{\Omega,\max}$ (\%) & $e_{\Omega,\mathrm{peak}}$ (\%) \\
\midrule
fp32   & 0.7198 & 3.7806 & 0.1764 \\
mxp-b1 & 0.6475 & \cellcolor{gray!20}\textbf{3.0603} & 0.2206 \\
mxp-b2 & 0.6777 & 3.4982 & 0.1313 \\
mxp-b3 & 0.6972 & 3.1276 & \cellcolor{gray!20}\textbf{0.1174} \\
mxp-b4 & 0.7783 & 3.7950 & 0.3661 \\
mxp-b5 & 0.7472 & 3.7123 & 0.1441 \\
mxp-b6 & \cellcolor{gray!20}\textbf{0.5221} & 6.9318 & 4.7076 \\
mxp-b7 & 0.7983 & 3.5525 & 0.2737 \\
\bottomrule
\end{tabular}%
}
\end{subtable}
\vspace*{-4mm}
\end{table}

\section{Conclusion and Future Work}
\label{sec:conclusion}
SEM-based CFD simulations such as the Taylor-Green vortex exhibit a different performance structure from solver-dominated linear algebra workloads. In the TGV cases studied here, the velocity and pressure solvers require only a small number of Krylov iterations per time step. As a result, the runtime is not dominated by Krylov convergence alone, but is distributed across the full $P_N-P_N$ fluid time-stepping loop. Motivated by this observation, we have proposed a three-level hierarchical mixed-precision control model for matrix-free SEM CFD simulations. The model exposes precision control at the block level, kernel level, and step level. 

We evaluated the block- and kernel-level controls using two groups of mixed-precision configurations, separated by the global precision bound. Group A, based on the rp=fp64 environment, mainly identifies accuracy-sensitive components and future optimization opportunities. In particular, SEM-focused fp32 compute appears to be a promising candidate once lower-precision GPU storage becomes fully resident and repeated precision conversions are reduced. Group B, based on the rp=fp32 environment, provides the main practical performance direction. For the high Reynolds number case considered, the best fp32-based mixed-precision candidate achieves approximately $34\%$ reduction in both time-to-solution and energy-to-solution while improving robustness compared with global single precision. We also explored the use of even lower precision through targeted fp16 kernel overrides. The fp16 results show that half precision can be useful in selected kernels.

Our future work will focus on three directions: 1/ extending the study to higher Reynolds number cases, where global single precision is expected to become much more sensitive; 2/ enabling step-based precision control to adapt the precision strategy during different stages of the simulation; 3/ developing more complete low-precision residency on GPUs to reduce conversion overhead and better expose the performance potential of mixed precision.

\subsubsection*{Acknowledgment} This research was partially supported by a Center of Excellence in Exascale CFD (CEEC) grant No 101093393 from \worldflag[width=3.2mm]{EU} via EuroHPC JU and \worldflag[width=3mm]{SE}\worldflag[width=3mm]{DE}\worldflag[width=3mm]{ES}\worldflag[width=3mm]{GR}\worldflag[width=3mm]{DK}.
%
We acknowledge usage of Karolina hosted by IT4I \worldflag[width=3mm]{CZ} via the EuroHPC EHPC-DEV-2025D02-126 development project. 

\vspace*{-2mm}

\bibliographystyle{splncs04}
\bibliography{references}

\end{document}